\documentstyle[prd,aps,twocolumn,amstex]{revtex}

\newcommand{\ket}[1]{|#1\rangle}
 
\newcommand{\nn}{\nonumber\\} 
\newcommand{\ul}{\underline} 
\newcommand{\f}[1]{\mbox{\boldmath$#1$}}

\newcommand{\bea}{\begin{eqnarray}}
\newcommand{\ea}{\end{eqnarray}}
\newcommand{\la}{\Lambda_{\rm QCD}}
\newcommand{\ord}{{\cal O}}

\begin{document} 

\wideabs{ 
\title{A cosmological constant from the QCD trace anomaly?}
\author{Ralf Sch\"utzhold} 
\address{
Department of Physics and Astronomy,
University of British Columbia,
Vancouver B.C., V6T 1Z1 Canada
\\
Electronic address: {\tt schuetz@@physics.ubc.ca}
}
\date{\today}

\maketitle

\begin{abstract} 
According to recent astrophysical observations the large scale mean pressure 
of our present universe is negative suggesting a positive cosmological 
constant like term. 
This article addresses the question of whether non-perturbative effects of
self-interacting quantum fields in curved space-times may yield a significant
contribution.
Focusing on the trace anomaly of quantum chromo-dynamics (QCD), a preliminary 
estimate of the expected order of magnitude yields a remarkable coincidence 
with the empirical data, indicating the potential relevance of this effect.
\\   
PACS: 
04.62.+v,  % Quantum field theory in curved spacetime
12.38.Aw,  % General properties of QCD (dynamics, confinement, etc.)
12.38.Lg,  % Other nonperturbative calculations (QCD)  
98.80.Es.  % Observational cosmology (including Hubble constant,
           % distance scale, cosmological constant, early Universe, etc)  
\end{abstract} 
}

Recent measurements of the cosmic microwave background 
\cite{boomerang} suggest that the large scale structure of our
universe is quite accurately described by the conformally flat
Friedmann-Robertson-Walker (FRW) metric 
\bea
\label{frw}
ds^2
=
d\tau^2-\Omega^2(\tau)\,d\f{r}^2
=
\Omega^2(t)\left(dt^2-d\f{r}^2\right)
\,,
\ea
with $\tau$ denoting the proper and $t$ the conformal time, respectively.
The temporal variation of the scale factor $\Omega^2$ inducing the
cosmological red-shift is represented by the Hubble parameter 
\bea
\label{hubble}
{\mathfrak H}=\frac{1}{\Omega}\frac{d\Omega}{d\tau}
\approx
10^{-10}{\rm year}^{-1}
\,.
\ea
Inserting the FRW metric in Eq.\ (\ref{frw}) into the Einstein equations
(with the cosmological constant $\lambda$) 
\bea
\label{einstein}
{\mathfrak R}_{\mu\nu}-\frac12{\mathfrak g}_{\mu\nu}{\mathfrak R}=\left(
8\pi\,G_{\rm N}\langle\hat T_{\mu\nu}\rangle - {\mathfrak g}_{\mu\nu}\lambda
\right)_{\rm ren}
\,,
\ea
the aforementioned observations and supernova \cite{supernova}
data consistently yield the following conclusions:
the $00$-component of the r.h.s.\ of Eq.\ (\ref{einstein}) equals 
(at least approximately) the critical density 
$\varrho\approx\varrho_{\rm crit}$ and the spatial $ii$-components -- 
associated with the pressure $p$ -- are negative: 
$p/\varrho_{\rm crit}\approx-2/3$. 
As a result the universe is presently undergoing an accelerated expansion as 
approximately described by the de~Sitter metric 
$\Omega(\tau)=\exp\{{\mathfrak H}\tau\}$.  

A negative pressure together with a positive energy density necessarily
implies a non-vanishing trace of the energy-momentum tensor $T_{\mu\nu}$.
Since $T_{\mu\nu}$ can be derived via the variation of the action $\cal A$
with respect to the metric ${\mathfrak g}^{\mu\nu}$, i.e.\
$T_{\mu\nu}=
2(-{\mathfrak g})^{-1/2}\delta{\cal A}/\delta{\mathfrak g}^{\mu\nu}$, 
its trace corresponds to the change of $\cal A$ under the conformal 
transformations 
${\mathfrak g}_{\mu\nu}(\ul x)\to\Omega^2(\ul x){\mathfrak g}_{\mu\nu}(\ul x)$
by virtue of Euler's law: 
$T^\rho_\rho=
-\Omega(-{\mathfrak g})^{-1/2}\delta{\cal A}/\delta\Omega$,
see e.g.\ \cite{birrell}.

Let us focus on the contribution of the $SU(3)$-color gauge field theory of
QCD to the r.h.s.\ of the Einstein equations (\ref{einstein}) in the following.
Its dynamics are governed by the well-known Lagrangian density 
\bea
\label{gauge}
{\cal L}=-\frac14\,G_{\mu\nu}^aG^{\mu\nu}_a
+\bar{\psi}\left(
i\gamma^\mu\partial_\mu+g\,T_a\gamma^\mu A^a_\mu-m
\right){\psi}
\,,
\ea
with $G_{\mu\nu}^a=\partial_\mu A_\nu^a-\partial_\nu
A_\mu^a+{g}f^a_{\;\;bc}A_\mu^b A_\nu^c$ 
being the gluonic field strength tensor.
Here $f^a_{\;\;bc}$ denote the $SU(3)$ structure constants,
$T_a$ its fundamental generators, and $g$ is the strong coupling.
For simplicity we drop the ghost fields as well as the gauge fixing terms
and consider only one single flavor, i.e.\ quark-species $\psi$.  
The remaining electro-weak sector of the standard model will be discussed
at the end of this article.

On the classical level all gauge field theories as described by
Eq.\ (\ref{gauge}) are conformally invariant (for $m=0$).  
According to the above arguments this feature implies a vanishing
trace of the classical energy-momentum tensor (as one can easily check
by an explicit calculation.) 
Turning to the quantum field theoretical description the situation
becomes more complicated.
In the first place, the na{\"\i}ve expectation value of the operator-valued
energy-momentum tensor diverges due to the infinite zero-point
energy. 
In order to renormalize this singularity by an appropriate counter-term one 
has to interpret the cosmological constant $\lambda$ in 
Eq.\ (\ref{einstein}) as a bare quantity \cite{birrell}. 

After such a minimal subtraction procedure the trace of the renormalized 
expectation value of the energy-momentum tensor 
$\langle\hat T_{\mu\nu}\rangle_{\rm ren}$ 
vanishes (again assuming $m=0$) for free (${g}=0$) fields in flat 
($\Omega=1$) space-times -- but {\em not} in the general case. 
This phenomenon is called the trace anomaly and
goes along with the dynamical breaking of the conformal
invariance of the classical theory in Eq.\ (\ref{gauge}).  
It has been calculated for two limiting cases: 
firstly for self-interacting quantum fields in flat ($\Omega=1$) 
space-times \cite{trace} and secondly for free ($g=0$) fields in curved
space-times \cite{birrell}.
In the first case one obtains \cite{trace}
\bea
\label{condensate}
\langle\hat T^\rho_\rho\rangle_{\rm ren}
=
\frac{\beta({g})}{2{g}} 
\langle\hat G_{\mu\nu}^a\hat G^{\mu\nu}_a\rangle_{\rm ren}
+
(1+\gamma^m)\,m\,\langle{\bar{\psi}}\hat\psi\rangle_{\rm ren}
\,.
\ea
The Callan-Symanzik or Gell-Mann--Low $\beta$-function describes the
scale dependence $\beta=\mu\,\partial{g}/\partial\mu$ of the renormalized 
coupling ${g}(\mu)$ and reflects the dynamical breaking of the conformal 
invariance of the classical theory (dimensional transmutation). 
Similarly, the $\gamma^m$-function corresponds to the running of the 
renormalized mass.
The expectation values
$\langle\hat G_{\mu\nu}^a\hat G^{\mu\nu}_a\rangle_{\rm ren}$ and 
$\langle{\bar{\psi}}\hat\psi\rangle_{\rm ren}$ 
occurring in Eq.\ (\ref{condensate})
represent the so-called gluonic and quark condensates, respectively,
see e.g.\ \cite{shifman}. 
These inherently non-perturbative quantities again reflect the dynamical 
breaking of the classical scale invariance.
Both are of great experimental relevance and their values have been 
confirmed within several contexts, see e.g.\ \cite{shifman}.
Since the symmetry breaking scale $\la$ is (for $m=0$) the only scale in 
the theory (\ref{gauge}), it yields 
$\langle\hat G_{\mu\nu}^a\hat G^{\mu\nu}_a\rangle_{\rm ren}
=\ord(\la^4)$ 
and 
$\langle{\bar{\psi}}\hat\psi\rangle_{\rm ren}=\ord(\la^3)$
-- at least for $m\ll\la$.
As it is well-known, the $\beta$-function occurring in  
Eq.\ (\ref{condensate}) can be calculated within the framework of 
perturbation theory and it turns out to be negative.
Consequently it is now commonly accepted (cf.\ \cite{shifman}) that
the QCD trace anomaly gives rise to a negative energy density
(since $\langle\hat G_{\mu\nu}^a\hat G^{\mu\nu}_a\rangle_{\rm ren}>0$
and $\langle{\bar{\psi}}\hat\psi\rangle_{\rm ren}<0$)
of the QCD vacuum in the Minkowski space-time.

However, such a huge amount of negative energy density of order 
$\ord(\la^4)$ blatantly contravenes our observations.
This drastic and global violation of the (weak and dominant) energy 
conditions (see e.g.\ \cite{birrell}) in the Minkowski space-time goes 
along with a fundamental contradiction if one includes gravity since the 
r.h.s.\ of the Einstein equations (\ref{einstein}) associated with 
a flat space-time vanishes. 

Consequently, regarding the Einstein equations (\ref{einstein}), one is led 
to absorb the aforementioned energy density by renormalizing the cosmological 
constant $\lambda$, in complete analogy with the case of the zero-point 
energy (which determines the divergent part of $\lambda$ only).
In the same manner as one adjusts the mass-counter term in the self-energy 
renormalization of the electron, for example, one has to fix the bare 
cosmological constant by demanding  that the r.h.s.\ of the Einstein 
equations (\ref{einstein}) vanishes for the Minkowski vacuum
\bea
\label{ren}
\left(
8\pi\,G_{\rm N}\langle\hat T_{\mu\nu}\rangle - {\mathfrak g}_{\mu\nu}\lambda
\right)_{\rm ren}^{\rm Minkowski\;vacuum}
=0
\,.
\ea

On the other hand, a non-trivial geometry of the space-time may also
induce a non-vanishing trace -- even for free fields 
(second limiting case $g=0$).
In this case $\langle \hat T^\rho_\rho\rangle_{\rm ren}$
is given by the sum of a bilinear form of the curvature tensor 
(such as ${\mathfrak R}_{\mu\nu}{\mathfrak R}^{\mu\nu}$ or 
${\mathfrak R}^2$)
and second derivatives of it ($\Box{\mathfrak R}$), cf.\ \cite{birrell}. 
For the free QCD field (with $g=m=0$) within the de~Sitter space-time
$\Omega(\tau)=\exp\{{\mathfrak H}\tau\}$, for example, one finds ($\hbar=c=1$)
\bea
\label{h^4}
\langle \hat T^\rho_\rho\rangle_{\rm ren}
=
%\frac{31}{15\pi^2}  % 0 quarks
%\frac{169}{80\pi^2} % 1 quarks
\frac{281}{120\pi^2} % 6 quarks
{\mathfrak H}^4
\,.
\ea
In contrast to the contribution in Eq.\ (\ref{condensate}), there is no 
reason to absorb this term by renormalization of $\lambda$.
In view of its potentially space-time dependent character such a 
procedure would be rather strange. 
However, here the associated energy density is far too small to 
explain the observations \cite{boomerang,supernova}.

In summary, the (renormalized) expectation value of the
energy-momentum tensor acquires an anomalous trace for self-interacting 
quantum fields in flat space-times (\ref{condensate}) on the one hand as well 
as for free fields in curved space-times (\ref{h^4}) on the other hand.  
However, both effects taken alone are not capable of explaining 
the negative pressure as suggested by the observations
\cite{boomerang,supernova}. 
But this is just what one might expect, since realistic 
investigations have to involve both contributions simultaneously, 
i.e.\ the (non-perturbative) effects of self-interacting fields in 
curved space-times. 
A rigorous derivation of the renormalized expectation value of the 
energy-momentum tensor for this scenario appears
to be rather involved and is not the aim of the present article.
Here we just give a preliminary estimate of the expected order of 
magnitude of the effect. 
To this end we employ an adiabatic approximation (cf.\ \cite{birrell})
by exploiting the huge difference of the involved time scales.
The cosmic evolution -- governed by ${\mathfrak H}$ -- is extremely slow 
compared to the typical fluctuations of the quantum field as determined by 
$\la$.
Consequently the adiabatic approximation is an expansion in the 
small parameter ${\mathfrak H}/\la=\ord(10^{-40})$. 
So the zeroth-order term is the pure Minkowski (flat space-time)
contribution whereas the first-order term represents the lowest
correction induced by the cosmic expansion. 

In order to calculate the renormalized expectation value of the 
energy-momentum tensor it is essential to specify the correct vacuum 
state associated with our expanding universe, cf.\ \cite{birrell}.
To this end we adopt the Schr{\"o}dinger picture
\bea
\frac{d}{dt}\ket{\Psi}=-i\hat H_{\rm\scriptstyle FRW}(t)\ket{\Psi}
\,,
\ea
where $\hat H_{\rm\scriptstyle FRW}(t)$ denotes the Hamilton operator, 
i.e.\ the generator of the time evolution, with respect to the conformal 
coordinates $(t,\f{r})$ in Eq.\ (\ref{frw}).
Within the adiabatic approximation, the explicitly time-dependent Hamiltonian 
$\hat H_{\rm\scriptstyle FRW}(t)$ of an expanding universe can be 
related to the (time-independent) Minkowski Hamiltonian 
$\hat H_{\rm\scriptstyle Min}$ via
\bea
\label{conformal}
\hat H_{\rm\scriptstyle FRW}(t)=
\exp\left\{-i\Omega(t)\hat S\right\}
\hat H_{\rm\scriptstyle Min}
\exp\left\{+i\Omega(t)\hat S\right\}
\,,
\ea
with $\hat S$ being the generator for the conformal transformations
${\mathfrak g}_{\mu\nu}\to\Omega^2{\mathfrak g}_{\mu\nu}$
in the Schr\"odinger picture.
In terms of a dynamically scaled state defined via
$\ket{\widetilde\Psi}=\exp\{+i\Omega(t)\hat S\}\ket{\Psi}$,
the Schr\"odinger equation assumes the form
\bea
\frac{d}{dt}\ket{\widetilde\Psi}=
-i\left(\hat H_{\rm\scriptstyle Min}-
\dot\Omega\hat S\right)\ket{\widetilde\Psi}
\,.
\ea
Treating $\hat H_1=-\dot\Omega\hat S=\ord({\mathfrak H})$ 
as a perturbation and switching to the interaction representation 
$\hat S(t)=\exp\{+i\hat H_{\rm\scriptstyle Min}t\}\,\hat S\,
\exp\{-i\hat H_{\rm\scriptstyle Min}t\}$
we may solve the above equation in linear response, i.e.\ first order 
adiabatic expansion
\bea
\label{response}
\ket{\widetilde\Psi}=\ket{\widetilde\Psi_{\rm in}}
+i\int\limits_{-\infty}^0 
dt\;\dot\Omega(t)\,\hat S(t)\,\ket{\widetilde\Psi_{\rm in}}
+\ord({\mathfrak H}^2)
\,.
\ea
If we assume $\Omega(t)=\Omega_0+\exp\{{\mathfrak H}t\}$, it is reasonable 
to take the Minkowski vacuum $\ket{0_{\rm\scriptstyle Min}}$ with
$\hat H_{\rm\scriptstyle Min}\ket{0_{\rm\scriptstyle Min}}=0$
as the initial condition 
$\ket{\widetilde\Psi_{\rm in}}=\ket{0_{\rm\scriptstyle Min}}$.
The remaining time integration yields 
$\hat H_{\rm\scriptstyle Min}^{-1}$ and hence we arrive at
\bea
\label{adiabatic}
{\ket{0_{\rm\scriptstyle FRW}}}=
{\ket{0_{\rm\scriptstyle Min}}}+
{{\mathfrak H}}\,\hat H_{\rm\scriptstyle Min}^{-1}\,\hat S\,
\ket{0_{\rm\scriptstyle Min}}
+\ord({\mathfrak H}^2)
\,.
\ea
Therefore the adiabatic QCD vacuum $\ket{0_{\rm\scriptstyle FRW}}$ 
of an expanding universe is {\em not} the instantaneous ground-state 
${\ket{0_{\rm\scriptstyle Min}}}$ of $\hat H_{\rm\scriptstyle FRW}(t)$ 
or $\hat H_{\rm\scriptstyle Min}$ -- it acquires corrections already in 
the first order of the adiabatic expansion. 
Instead it is the ground-state of the corrected Hamiltonian 
$\hat H_{\rm\scriptstyle Min}-\dot\Omega\hat S$, 
which can easily be verified using stationary perturbation theory
(see e.g.\ \cite{birrell} for free fields).

The remaining question is, of course, whether the first-order correction 
${\mathfrak H}\,\hat H_{\rm\scriptstyle Min}^{-1}\,
\hat S\,\ket{0_{\rm\scriptstyle Min}}$ to the vacuum state 
entails a first-order correction to the expectation value of
$\hat T_{\mu\nu}$.
In order to illustrate this point let us consider the simple example of a
time-dependent harmonic oscillator
\bea
\label{free}
\hat H(t)=\frac{\omega}{2\Omega(t)}
\left(\hat P^2+\Omega^2(t)\hat Q^2\right)
\,,
\ea
where $\omega$ corresponds to $\la$.
In examining the question of whether the dynamical scale symmetry 
breakdown in QCD can be 
modeled by such a simple quadratic potential one might consider the
$CP(N-1)$ or the $O(N)$ $\sigma$-models \cite{cpn}.
These strongly interacting theories reproduce several features
of QCD, such as dynamical scale symmetry breakdown.
They can be solved in the large $N$-limit and in the leading order they 
effectively behave like massive free fields. 
After a normal mode decomposition one therefore indeed obtains 
terms like the one above.    
For the Hamiltonian in Eq.\ (\ref{free}) a change of the scale factor 
$\Omega(t)$ as in Eq.\ (\ref{conformal}) can simply be generated by the 
squeezing operator
\bea
\label{squeezing}
\hat S=\frac{1}{4}\left\{\hat P,\hat Q\right\}=
\frac{i}{4}\left[\left(\hat a^\dagger\right)^2
-\left(\hat a\right)^2\right]
\,.
\ea
Consequently the expectation values of operators such as $\hat Q^2$ or
$\hat P^2$ do not acquire a first-order correction.
This result can be transfered directly to free quantum fields:
pitching on a particular normal mode with the wavelength $k$ the conformal 
charge $\hat S$ again acts like a squeezing operator $\hat S\to 
i(\hat a_{k}^\dagger\hat a_{-k}^\dagger-\hat a_{k}\hat a_{-k})$.
As a result there is no first-order correction to the expectation value of
$\hat T_{\mu\nu}$ for free (linear) fields.

However, if we leave the free-field sector and take interactions into
account the situation may change:
let us consider the interaction Hamiltonian
\bea
\label{inter}
\hat H_{\rm int}(t)=g
\int\limits_{-\infty}^{t} dt'\,G_{\rm ret}(t-t')
\left\{\hat Q^2(t),\hat Q^2(t')\right\}
\,,
\ea
where $\hat Q(t)=\hat Q\cos(\omega t)+\hat P\sin(\omega t)/\omega$ denotes 
the unperturbed time-dependent operator in the interaction picture.
The retarded propagator $G_{\rm ret}(t-t')$ encodes the dynamics of an 
intermediate (interaction) degree of freedom which has been integrated out.
Again such a term can be motivated by the $CP(N-1)$-models:
in the large $N$-limit these strongly interacting massless theories
effectively transform into massive fields obeying weak 
(i.e.\ next-to-leading order in $1/N$) long-range four-point interactions, 
cf.\ \cite{cpn}.

For general Green functions $G_{\rm ret}(t-t')$,
the operator in Eq.\ (\ref{inter}) does entail a first-order correction. 
This can be most easily verified by assuming $g\ll1$ which allows for a
perturbative treatment. 

In view of these considerations one might expect a first-order contribution 
to $\langle\hat T_{\mu\nu}\rangle_{\rm ren}$ to be possible in the case 
of QCD:
since the classical as well as the free quantum field in Eq.\ (\ref{gauge})
are (for $m=0$) conformally invariant, their solutions would simply be 
scaled during the expansion of the universe -- like the red-shift of the
photon field.
(This would actually happen if the universe were to expand very rapidly
${\mathfrak H}\gg\la$.)
However, the strong self-interaction on the quantum level breaks the 
conformal invariance and introduces a fixed scale $\la$ leading to a 
positive pressure given by Eq.\ (\ref{condensate}).
Within an expanding universe the balance of these two tendencies,
i.e.\ following the expansion on the one hand and retaining the scale
on the other hand, leads to a displaced vacuum state (\ref{adiabatic}).

Let us assume that a part of the positive vacuum pressure in 
Eq.\ (\ref{condensate}) can be explained by relatively localized 
(non-perturbative) vacuum fluctuations 
(e.g.\ instantons \cite{instanton} or oscillons \cite{oscillon}) 
which repel each other (at least in average, cf.\ \cite{instanton}).
Let us further assume that the dynamical breakdown of the scale symmetry
is basically encoded by these (non-perturbative) vacuum-fluctuations whereas
their (repulsive) interactions are adequately described by the free 
(perturbative) and thus conformally invariant field equations.
In this case their solutions would simply be scaled during the cosmic 
expansion in contrast to the non-perturbative fluctuations which retain 
their scale and hence are not affected. 
Within an expanding universe, then, every vacuum fluctuation ``sees''
all other vacuum-fluctuations ``red-shifted'', i.e.\ their repulsion
acquires a correction proportional to ${\mathfrak H}R$, 
where $R$ denotes their (mean) distance (cf.\ \cite{instanton}). 
Accordingly, the positive vacuum pressure in Eq.\ (\ref{condensate}) 
gets diminished by an amount of first order in ${\mathfrak H}$.

Based on this intuitive picture it appears plausible to admit a correction
to the expectation value of the energy-momentum tensor within the FRW vacuum 
in Eq.\ (\ref{adiabatic}) which is linear in ${\mathfrak H}$.
After the renormalization described in Eq.\ (\ref{ren}), i.e.\ the 
subtraction of the Minkowski contribution, we therefore obtain
\bea
\label{result}
\langle\hat T_{\mu\nu}\rangle^{\rm FRW}_{\rm ren}
&=&
{\mathfrak H}
\langle\,\hat T_{\mu\nu}\,\hat H_{\rm\scriptstyle Min}^{-1}\,
\hat S\,\rangle^{\rm Min}_{\rm ren}
+{\rm H.c.}+\ord({\mathfrak H}^2)
\nn
&=&
\ord({\mathfrak H}\la^3)
\,.
\ea
Let us estimate the associated order of magnitude:  
although $\la$ depends on the renormalization scheme we may fix it 
approximately via $\la=\ord(10^8\,{\rm eV})=\ord(10^{14}\,{\rm m}^{-1})$.
The masses of the light quarks which dominantly couple to the gluonic field
are roughly of a similar order of magnitude.
The Hubble expansion parameter ${\mathfrak H}$ is about 
$10^{-26}\,{\rm m}^{-1}$.
Inserting the above values we finally arrive at 
$\langle\hat T_{\mu\nu}\rangle_{\rm ren}^{\rm FRW}=
\ord(10^{16}\,{\rm m}^{-4})$
or
$\langle\hat T_{\mu\nu}\rangle_{\rm ren}^{\rm FRW}=
\ord(10^{-29}\,{\rm g}\,{\rm cm}^{-3})$.
By inspection one finds that the deduced order of magnitude 
nicely fits the empirical data
$\varrho_{\rm crit}\approx10^{-29}\,{\rm g}{\rm cm}^{-3}$.
In view of the huge difference of the involved scales 
[${\mathfrak H}/\la=\ord(10^{-40})$] this remarkable coincidence seems 
to be almost too good to be just an accident.
At least it indicates the potential relevance of the effect described
in the present article with regard to the interpretation of the
astrophysical data \cite{boomerang,supernova}.  

It should be mentioned here that a pressure induced by the expansion of 
our universe with $p\propto{\mathfrak H}$ generates a cosmic evolution 
which differs from that with a true cosmological constant $p=\rm const$:
by inserting the FRW metric (\ref{frw}) into the Einstein equations 
(\ref{einstein}) one obtains the Friedmann equation 
$3{\mathfrak H}^2=8\pi\,G_{\rm N}\varrho$.
Furthermore the Einstein equations imply 
$\nabla_\mu\langle\hat T^{\mu\nu}\rangle_{\rm ren}=0$, 
i.e.\ $d\varrho/dt=-3(\varrho+p){\mathfrak H}$.
Combining these two equalities and specifying the pressure $p$ one may
determine the time-evolution of our universe.
Unfortunately the presently available data 
(such as the Hubble parameter or the age of the universe)
are not precise enough to distinguish the two cases
($p\propto{\mathfrak H}$ and $p=\rm const$).

Of course one may ask whether the remaining electro-weak sector of the 
standard model generates similar contributions: 
typically (see e.g.\ \cite{shifman}) non-perturbative effects 
(such as $\langle\hat G_{\mu\nu}^a\hat G^{\mu\nu}_a\rangle_{\rm ren}$ and 
$\langle{\bar{\psi}}\hat\psi\rangle_{\rm ren}$) display a 
dependence on the coupling of $\exp\{-8\pi^2/g^2\}=\exp\{-2\pi/\alpha\}$.
The scale of the dynamical symmetry breaking $\la$ obeys a similar
non-analytical dependence on the coupling $g$.
Inserting $\alpha_{\rm QED}\approx1/137$ into the above expression one
obtains a suppression by an order of magnitude of $10^{-370}$.
Hence the contributions arising from the dynamical breaking of scale 
invariance can safely be neglected in this case.

The remaining explicit breaking of the scale symmetry induced by the Higgs
field of course also generates contributions to $\hat T_{\mu\nu}$.
However, the general structure of all these terms is given by
$m^2\,\langle\hat\Phi^\dagger\hat\Phi\rangle_{\rm ren}$
and according to the arguments after Eq.\ (\ref{squeezing}) 
they do not contribute to the first order in ${\mathfrak H}$.
Although the mixture of these terms caused by interactions remains 
subject to further considerations, a contribution of the electro-weak 
sector in analogy to QCD is not obvious.

In summary, the present article motivates a deeper examination of the vacuum 
of strongly interacting fields in the gravitational background of our 
expanding universe 
-- for the present epoch as well as for earlier stages, cf.\ \cite{inflation}.
These investigations might perhaps lead to a better understanding of some 
of the problems in cosmology without necessarily invoking yet unknown 
low-energy fields, for example quintessence (see e.g.\ \cite{quint}).

\newpage

%{\em Acknowledgments}
%\\
The author is indebted to B.~Unruh and E.~Zhitnitsky for valuable discussions.
This work was supported by the Alexander-von-Humboldt foundation and by NSERC.

\addcontentsline{toc}{section}{References}%%%%%%%%%%%%%%%%%%%%%%%%%%%%%%%%

\end{document}